\documentclass[twocolumn,aps,prb,floatfix,superscriptaddress,showpacs]{revtex4}
\usepackage{graphicx}
\usepackage{bm}
\usepackage{epsfig}

\newcommand{\dsas}{\Delta_\mathrm{SAS}}
\newcommand{\dsasr}{\tilde\Delta_\mathrm{SAS}}
\newcommand{\dsw}{\Delta_\mathrm{sw}}

\begin{document}
\title{Quantum and Classical Dissipative Effects on
Tunnelling in Quantum Hall Bilayers}

\author{Robert L. Jack}
\affiliation{Theoretical Physics, 1 Keble Road, Oxford OX1 3NP, United
Kingdom}
\author{Derek K. K. Lee}
\affiliation{Blackett Laboratory, Imperial College, Prince Consort Road,
 London SW7 2BW, United Kingdom}
\author{Nigel R. Cooper}
\affiliation{Cavendish Laboratory, Madingley Road, Cambridge CB3 0HE,
United Kingdom}

\begin{abstract}
We discuss the interplay between transport and dissipation in quantum
Hall bilayers. We show that quantum effects are relevant in the pseudospin
picture of these systems, leading either to direct tunnelling currents
or to quantum dissipative processes that damp oscillations around the
ground state. These quantum effects have their origins in resonances of
the classical spin system.
\end{abstract}
\maketitle



Over the past ten years, a series of elegant 
experiments\cite{Eisenstein92-94,Suen1992,Spielman2000on,Kellogg2004,Tutuc2004}
on quantum Hall
bilayer systems have allowed investigation of the correlated many-body
quantum state at total filling fraction $\nu=1$. 
The bilayers consist of two closely-spaced parallel 
two-dimensional electron layers in a double quantum well. If the Landau
level fillings are $\nu_1=\nu_2=1/2$ then separate layers will not
show the quantum Hall effect. However, Coulomb interactions between the
layers drive 
a transition into a state in which the bilayer as a whole
exhibits quantised Hall conductance\cite{qh_pers}. 
Predictions\cite{bl_json} that
these states would 
undergo spontaneous breaking of a $\mathrm{U}(1)$ symmetry have been 
supported by experiments\cite{Eisenstein92-94}. 

The ground state of the system may be viewed as an easy-plane
ferromagnet\cite{Moon1995}.
There also are analogies with 
Josephson junctions\cite{bl_json} and excitonic 
superfluids\cite{Fertig1989,Balents2001}.
The transport properties of 
the bilayers are rich and often surprising. They provide experimental
evidence for the Goldstone mode associated with the broken 
$\mathrm{U}(1)$ symmetry\cite{Spielman2000on} and, more recently,
`excitonic superfluidity'\cite{Kellogg2004,Tutuc2004}. 

The interlayer tunnelling properties of the bilayer remain a subject
of considerable theoretical
study\cite{bl_theory,Abolfath2004,Bezuglyj2004,Fertig2003,Wang2004,Park2004cm}.
In a previous
paper\cite{Jack2004} we introduced a simple thought experiment in
which we investigated the dissipation and interlayer tunnelling
properties of a clean bilayer system at zero temperature, and gave a
brief comparison of our theoretical predictions with the experimental
results. A key point is the existence of a 
crossover\cite{Abolfath2004,Bezuglyj2004} in
behaviour at a very small bias $V_0$ which corresponds to an energy
of the order of the energy gap of the collective charge
excitations in the bilayer. This gap is small for weak tunnelling and
experiments operate principally at $V>V_0$. As mentioned in
Ref.~\onlinecite{Jack2004}, we predict a tunnelling current that
varies as $1/V$ in that region. This negative differential conductance
is in qualitative agreement with experiment.

In this paper, we extend the discussion of Ref.~\onlinecite{Jack2004},
including details of the behaviour for $V<V_0$, and a discussion
of the relationship between quantum dissipative processes and instabilities
of a classical spin system.
Section~\ref{sec:model} contains a discussion of 
the model and of our thought experiment. In section~\ref{sec:qu} 
we discuss the results both 
for $V>V_0$, and for $V<V_0$. In the latter regime, the theoretical treatment
is made rather complicated by the presence of an out of equilibrium Bose 
condensate 
in the low energy bosonic theory of the bilayer: we show how these issues
may be resolved.
In section~\ref{sec:cl}, 
we show how the quantum dissipative mechanisms in 
the system arise from instabilities of an underlying classical spin system. 
Finally we draw conclusions in section~\ref{sec:conc}, and identify
outstanding issues in our treatment.

\section{The Model}
\label{sec:model}

In this section, we define our model, and describe the thought experiment 
that we will
use to investigate the link between dissipation and tunnelling.
The bilayer system may be cast in a pseudospin formulation\cite{Moon1995}.
The layer index of each electron defines a two dimensional Hilbert space: it
resembles the usual spin-(1/2) degree of freedom of electrons. We assume
throughout that physical electron spin is completely polarised by the applied
magnetic field. In the quantum Hall state, we may work in a basis of localised
orthogonal states in the lowest Landau level. 
Configurations in which two electrons occupy the same localised state represent
quasiparticles, and have a large energy gap.  Therefore,
a theory of discrete spins ($S=1/2$) on a lattice is appropriate at low 
energies.  The spins interact by an exchange
interaction with an easy plane, and experience a field in the easy plane
that arises from tunnelling between the layers.

In our calculations, we generalise the model to $S>(1/2)$: this
generalisation may be treated as a coarse-graining procedure.  The
limit of large $S$ is also the classical limit for the spin system.
After coarse-graining, we replace the easy-plane exchange interaction
by a combination of isotropic exchange and an on-site anisotropy term.
The resulting Hamiltonian is:
\begin{equation}
\frac{H}{2S} = - \frac{\rho_E}{2} \sum_{\langle ij \rangle} \vec{m}_i
\cdot \vec{m}_j + \frac{D}{4} \sum_i ( m_i^z )^2 -
\frac{\Delta_\mathrm{SAS}}{2} \sum_i m^x_i
\label{equ:H_spins}
\end{equation}
where $\vec{S}_i=S\vec{m}_i=S(m_i^x,m_i^y,m_i^z)$ is the spin operator
on site $i$ of a square lattice with spacing $c_0=\sqrt{2\pi} l_B$
where $l_B=(\hbar c/eB)^{1/2}$ is the magnetic length.  The interlayer
exchange $\rho_E$ and the strength of the on-site repulsion
$D$ were derived from microscopic considerations by Moon \emph{et
al.}\cite{Moon1995} ($D=8\pi\beta l_B^2$ in the notation of that
paper). 
The tunnelling between the layers enters the problem through
$\dsas$: the splitting of the ``bonding'' and ``anti-bonding''
single-particle states in the double well.
 We use a Hamiltonian with isotropic exchange purely for simplicity: the
 interlayer and intralayer exchange constants will differ in general.
 However, if we generalise to an exchange interaction that is anisotropic
 (in the spin space), then only the interlayer part is relevant at the
 large lengthscales and small charge imbalances that are relevant to our 
 calculations 
 (we require $m^z\ll 1$, and
 $\rho_E q^2 \ll D$ for all relevant wavevectors, $q$).
 Since these conditions are satisfied for the experimental comparisons
 that we make, we set both the interlayer and intralayer exchange
 constants to their interlayer value, $\rho_E$.
In our thought experiment, a gate is used to control
the charge imbalance on the bilayer: this adds a term $H_V = - SV
\sum_i m^z_i$ to the Hamiltonian.

Typical values for the model parameters in physical bilayer systems
are $l_B\simeq20\mathrm{nm}$,
$\dsas\simeq90\mu\mathrm{K}$, $\rho_E\simeq0.5\mathrm{K}$, $D\simeq
30\mathrm{K}$.

Hamiltonians of the form (\ref{equ:H_spins}) permit both ferromagnetic and
quantum disordered phases at zero temperature. At large $D$
(or small $S$), the uncertainty in $S^z$ associated with ferromagnetic order 
means that the paramagnetic phase with $S^z\to0$ becomes energetically
favoured. We interpret the observation of a linearly dispersing peak in the 
tunnelling conductance\cite{Spielman2000on} as evidence that the
experimentally accessible states have ferromagnetic order 
--- the peak arises from the Goldstone mode of the system.
In other words, we assume that the system is connected adiabatically
to the large-$S$ limit of our model, ignoring the possibility
of the quantum disordered phase that exists for $D/\rho_E S^2 \gg 1$
\cite{Chubukov1994}.

We seek a minimal theory of the bilayer. We therefore
work in a clean system at $T=0$. This allows us to calculate dissipation
rates starting from purely microscopic considerations: this is distinct
from the more phenomenogical approach of Ref.~\onlinecite{Abolfath2004}
in which dissipation is added to the model in the form of damping terms 
in the classical equations of motion.
Simulations\cite{Fertig2003} indicate
that disorder may be relevant to the tunnelling at small bias voltages.
While these effects are beyond the scope of this paper, 
we argue below that our framework may be suitable for further investigation 
of these effects. Possible effects of topological defects such as
merons\cite{qh_pers} are also neglected in our treatment, 
as are inhomogeneities in the order parameter associated with
the spontaneous symmetry breaking\cite{Wang2004}. 

A feature of our
framework is that we work on transport in an isolated bilayer. This
means that our calculations are well-controlled. Coupling
the bilayer to leads, as in experiments, complicates the picture. We
believe that the bulk effects that we study should dominate over
boundary effects associated with the leads, such as those studied 
by Park\cite{Park2004cm}, especially in the (experimentally relevant)
geometry where the current source and drain are separated by a distance 
of the order of the sample size.

Having reduced the system to a minimal model, we now proceed to 
discuss the behaviour of the spin system of
equation~(\ref{equ:H_spins}).


\subsection{Classical limit of the spin system}
\label{sec:cl_spins}

\begin{figure}
\epsfig{width=0.8\columnwidth,file=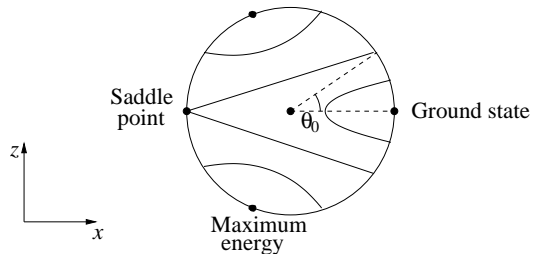}
\caption{Sketch of the constant energy trajectories on the spin sphere. }
\label{fig:q0_spins}
\end{figure}

In the ferromagnetic phase, we expect the qualitative behaviour of the
system to be accessible through an expansion about the classical limit.
In this limit, an initially uniform configuration of the spins remains 
uniform forever. Each spin precesses along one of the trajectories of 
constant energy shown in figure~\ref{fig:q0_spins}. There are two
qualitatively different types of trajectory in that figure: those which
wind around the $x$-axis and those which wind around the $z$-axis.
This distinction underlies the crossover in the tunnelling $IV$ relation
mentioned above. We now discuss the form of the trajectories 
and the classical properties of the spin system.

Trajectories with low energies are localised near the ground state, 
in which the spins are aligned along the $x$
axis. The spins precess around that axis with
a frequency $\dsw/\hbar$ where
\begin{equation}
\dsw=[\dsas ( \dsas + D )]^{1/2}
\label{equ:swgap}
\end{equation}
is the energy gap for spinwave
excitations. These excitations describe density waves of charge imbalance
across the bilayer.  For the case $\dsas=0$ these excitations
form the Goldstone mode that appears when the symmetry around the 
easy plane is spontaneously broken. In that case they disperse linearly
with a velocity $v=l_B (2\pi D\rho_E)^{1/2}$. 
Introducing the tunnelling $\dsas$ breaks the 
symmetry explicitly, and the gap opens up.
As the mean magnetisation precesses around the $x$-axis,
the charge imbalance on the bilayer oscillates around
zero.  

For trajectories with much larger energies,
the mean spin precesses around the
maximal energy state, which is close to the $z$-axis of the spin
sphere. This yields a Josephson-like alternating current $I\simeq
e\dsas \cos (eVt/\hbar)$ where $V$ is the voltage across the bilayer
due to capacitative charging.  We stress that this is valid only for
large charge imbalance (large $V$).

A saddle point on the negative $x$-axis of the spin sphere
marks the boundary between oscillations around the ground state and
oscillations around the maximal energy state.
The saddle point trajectory crosses the  $xz$-plane
at the saddle point, and at $(x,z)=(S\cos\theta_0,S\sin\theta_0)$ where
$\theta_0$ is the angle shown in Fig.~\ref{fig:q0_spins}. It is given by
$\cos\theta_0=1-2(\dsas/D)$. 

When the gate is used to induce a charge
imbalance on the bilayer, the mean spin is tilted out of the easy
plane by an angle $\theta$, which satisfies:
\begin{equation}
eV = D\sin\theta + \dsas\tan\theta 
\label{equ:V_theta}
\end{equation} 
where $V$ is the gate voltage.
The saddle point then corresponds to a voltage difference of
\begin{equation}
V_0 = (2\dsw/e) [ 1+\mathcal{O}(\dsas/D) ]
\end{equation}
across the layers. (We write $V_0$ in this form since 
the tunnelling energy scale set by $\dsas$, is much smaller
than the charging energy scale set by $D$.)
Since the dynamics of this
system depends strongly on whether the applied bias is above or below
this threshold value, the calculations we describe below will treat 
these two regimes separately.


\subsection{A thought experiment}

We now turn to our thought experiment which will form the basic
strategy behind our calculations. We imagine using a gate to induce a
uniform charge imbalance on the bilayer. In the spin picture, the
magnetisation is tilted out of the easy ($xy$) plane with 
\begin{equation}
m^z = \sin\theta \simeq eV/D
\label{equ:tiltangle}
\end{equation}
where $V$ is the voltage bias applied across the bilayer and the
second approximate equality is valid to leading order in $(\dsas/D)$
and $(eV/D)$, both of which will be small (recall equation
(\ref{equ:V_theta})).  The
magnetisation remains in the $xz$-plane due to the tunnelling
field. The bias is then instantaneously removed and the bilayer finds
itself in a highly excited state. In this paper, we will calculate the
quantum and classical dynamics of this highly excited system.

\begin{figure}
\epsfig{width=0.8\columnwidth,file=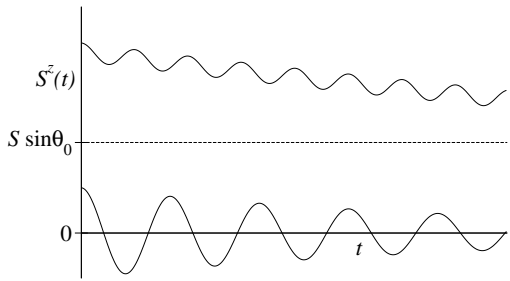}
\caption{Sketch of the charge imbalance on the bilayer ($S^z$)
in the thought experiment,
as a function of time after the gate voltage is removed.}
\label{fig:sz_t}
\end{figure}

From the discussion of the classical dynamics,
we expect the thought experiment to give different results
according to whether the initial gate voltage $V$ is above or
below the saddle point value $V_0$. For small initial energy, $V<V_0$,
the charge imbalance oscillates around zero: the effect of any dissipation 
will be to damp this oscillation (lower trace in figure~\ref{fig:sz_t}).  
Conversely, if the initial charge is larger, the
oscillation is about a finite charge imbalance: dissipation leads to
a decay of that charge imbalance, and therefore a direct current (upper
trace in figure~\ref{fig:sz_t}). The link between dissipation and
transport is clear in this case, since dissipative processes lead directly
to a direct current.

In our calculations,
we will treat these two regimes in separate perturbation theories.
Common to both calculations, we will expand around
the classical limit using spinwave
theory in a $1/S$ expansion. In a quantum theory of the spin system, 
the coherent oscillations of the charge imbalance do not remain
coherent forever. This is different from the classical case and is
consistent with the mixing between modes with different wavevectors observed 
in the exact diagonalisations of Nakajima\cite{Nakajima2001}. Dissipation
is possible because the tunnelling term breaks the global spin rotation
symmetry so that the long-wavelength modes are no longer protected from
decay by Goldstone's theorem.
We will find that the leading terms in
the expansion about the classical limit 
result in the leading terms in the relations between
the dissipation rate and the initial gate voltage, $V$.

\section{Dissipation rates in the quantum system}
\label{sec:qu}

In this section, we calculate the dissipation rates out of the coherent
classical oscillation of the spin system. As shown in figure~\ref{fig:sz_t}, 
the effect
of the dissipation is different, depending on the initial charge on the bilayer.
However, in both cases, the dissipation rate, $\Gamma$ is given by
\begin{equation}
\Gamma= \partial_t \Big< i \Big| e^{iHt} \sum_{\bm{q}\neq0}
E_{\bm{q}} e^{-iHt}
\Big| i \Big>
\label{equ:Gamma_def}
\end{equation}
where $E_{\bm{q}}$ measures the energy in the bosonic mode with momentum 
$\bm{q}$ and $|i\rangle$ is an initial state for the system.
The procedure used to calculate the dissipation rate is different 
in each case.

The smallest energy scale in the physical bilayer systems is $\dsas$, so it
seems natural to treat this parameter perturbatively. However, the 
tunnelling term breaks a continuous symmetry of the Hamiltonian, so
care must be taken in the
limit $\dsas\to0$.  It is clear from figure~\ref{fig:q0_spins} 
that the trajectories with 
energies above the saddle point energy are perturbatively connected
to the $\dsas=0$ limit. Conversely, trajectories with very low energies
are qualitatively different to those in the limit of zero tunnelling: 
in this case perturbation theory in $\dsas$ is not appropriate. Instead,
we treat the energy of the trajectory (or equivalently the initial gate 
voltage) perturbatively.

We now calculate the extent to which
quantum dissipative processes affect charge relaxation on the bilayer,
or damping of charge oscillations (recall figure~\ref{fig:sz_t}).

\subsection{Charge relaxation: $V>V_0$}
\label{sec:qu_asp}


We begin by considering the situation for initial energies above the saddle
point ($V>V_0$). This calculation was outlined in Ref.~\onlinecite{Jack2004}; 
we review the 
method here since the calculation in the following subsection 
is a refinement of the same approach. Evaluation of $\Gamma$ requires
the choice of an initial state, and calculation of matrix elements
between that state and
final states with different values of $E_{\bm{q}}$. We use
a bosonic representation of the spin algebra, which defines the basis
for initial and final states. We then identify the initial state
and the dominant process that leads to dissipation; this allows calculation
of $\Gamma$. 

To begin, we define our bosonic representation of the Hamiltonian, by
using the Holstein-Primakov representation of the spin algebra
$S^x_j+iS^y_j=(2S-a^\dag_j
a_j)^{1/2} a_j$, $S^z_j = S-a^\dag_j a_j$.  
We treat tunnelling perturbatively, so we begin by setting
$\dsas=0$ and expanding around
$m^z=(eV/D)=\sin\theta$. The
quadratic part of the Hamiltonian is then easily diagonalised 
in the Fourier basis to give
\begin{equation}
H^{(0)}_{\theta>\theta_0} = (D\sin\theta) \delta n_{q=0}
  + \sum_{\bm{q}\neq0} \varepsilon_{\bm{q}} \alpha_{\bm{q}}^\dag \alpha_{\bm{q}}
\end{equation}
where the $\alpha_{\bm{q}}$ are bosonic operators
($[\alpha_{\bm{q}},\alpha^\dag_{\bm{q'}}]=\delta_{\bm{q},\bm{q'}}$)
describing the spinwave modes. The spinwave dispersion is given by
\begin{eqnarray*}
\varepsilon_{\bm{q}} &=& [\rho_E \gamma(\bm{q}) ( D + \rho_E
\gamma(\bm{q}))]^{1/2}\\
\gamma(\bm{q})&=&4-2\cos(q_x c_0) - 2\cos(q_y c_0).
\label{equ:swhighbias}
\end{eqnarray*}
As discussed above, the tunnelling is being treated perturbatively.
Therefore, this dispersion relation is that of a model with spin
rotational symmetry in the $xy$-plane, and $\varepsilon_{\bm{q}}$ is
gapless linear in $q$ at long wavelengths.  The absence of an energy
gap in this approximation does not affect our calculation for $V>V_0=2\dsw/e$.
Observe that the $q=0$ mode has been singled out in the Hamiltonian,
and its energy is not given by the long wavelength limit of
$\varepsilon_{\bm{q}}$.  The quanta of this mode carry $S^z=1$ while
spin waves with finite wavevector have $S^z=0$. Thus the eigenspectrum
of the system consists of multiple branches: each branch has an
associated $S^z$, and consists of a continuum of linearly dispersing
collective modes. We use a basis $|N,\{\bm{q},\bm{q'},\dots\}\rangle$
where $N$ is the number of quanta in the $q=0$ mode (which labels the
branch), and the set of wavevectors indicate indicate the presence of
collective modes with finite momenta.

The next step is to express the initial state of the system in this basis.
The thought experiment stipulates an initial state with the expectation of
the spin equal to $(S\cos\theta,0,S\sin\theta)$: this is a coherent state.
However, a choice of an eigenstate of $S^z$ with the same expectation of
the energy leads to the same average rate. Thus we use
$|i\rangle_{V>V_0} = | (eV/D) (S L^2) , \{\} \rangle$.

\begin{figure}
\epsfig{file=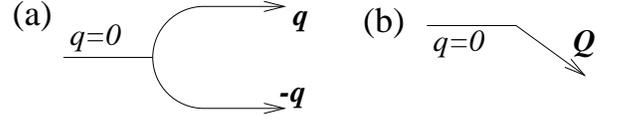,width=0.9\columnwidth}
\caption{Decay processes above the saddle point: (a) in the
absence of an in-plane magnetic field, we see decay of a single
quantum of the $q=0$ mode into a spinwave pair that satisfies
$\varepsilon_{\bm{q}} = \varepsilon_{-\bm{q}} = eV/2$. (b) 
in the presence of a in-plane field, a single quantum
of the $q=0$ mode can decay into a spinwave excitation with
momentum $Q$.}
\label{fig:diag_asp}
\end{figure}

The final step in the calculation
is to identify the dominant process contributing to $\Gamma$.
The dissipation arises from the destruction of one quantum in the
$q=0$ mode, and the
generation of multiple spinwaves during tunnelling across the
bilayer.  To leading order in $1/S$, a pair of spin waves is excited,
each with energy $eV/2$ but with opposite momenta
(Fig.~\ref{fig:diag_asp}a).  The relevant vertex in the interaction
Hamiltonian is:
\begin{equation}
H^{(1)}_{\theta>\theta_0} = -\frac{\dsasr}{8} \left[
e^{-i\phi_{q=0}}
\sum_{\bm{q}} \gamma_{2\bm{q}} \alpha^\dag_{\bm{q}} \alpha^\dag_{-\bm{q}}  +
\hbox{h.c.}  \right]
\end{equation}
in which $\dsasr = \dsas \exp(-S^{-1}\sqrt{D/\rho_E})$ is a
renormalised tunnelling amplitude, and the vertex factor is given by
$\gamma_{2\bm{q}} = \cos\theta [ (u_{\bm{q}}+v_{\bm{q}})^2 +
2\sin\theta\sec^2\theta- (u_{\bm{q}}-v_{\bm{q}})^2\sec^4\theta ] $
where $u_{\bm{q}}$ and $v_{\bm{q}}$ are coherence factors:
$(u_{\bm{q}}+v_{\bm{q}})^2=(D+\rho_E\gamma(\bm{q}))/\varepsilon_{\bm{q}}$
with $u_{\bm{q}}^2- v_{\bm{q}}^2 =1$. The $\theta$-dependence arises
because the $x$-component of the pseudospin depends on the angle
$\theta$ of the spin with the easy plane, as well as the azimuthal
angle $\phi$ around the plane. This dependence is weak for
$\theta\ll 1$ when the charge imbalance is small
compared to the Landau level filling.

Hence we calculate the power dissipation
$\Gamma$ for a given initial voltage $V$, using 
$E_{\bm{q}}=\varepsilon_{\bm{q}}
\alpha^\dag_{\bm{q}} \alpha_{\bm{q}}$. 
The calculation proceeds according to Fermi's Golden Rule.
The steady-state tunnelling
current density at a bias $V$ can then be computed from this dissipation by
$I=\Gamma/VL^2 S$. We find:
\begin{equation}
\frac{\Gamma_{\theta>\theta_0}}{L^2 S} = I_{\theta>\theta_0}V =
\frac{D \dsasr^2}{32 \pi \rho_E l_B^2\hbar S}
\left[ 1 + X(\theta) \right]^2
\label{equ:I_asp}
\end{equation}
where $X(\theta)=(\sec\theta-1)(\sec^2\theta-2\sec\theta-1)$ is small
for a small charge imbalance ($\theta\ll 1$). The dissipation saturates
at small initial gate voltages, leading rise to an increasing current
as the gate voltage is reduced.

As discussed in Ref.~\onlinecite{Jack2004}, 
this form for the current is qualitatively
consistent with the experiments of Ref.~\onlinecite{Spielman2000on}, 
in that the region of
negative differential conductance is observed for $V>V_0\sim 6\mu\mathrm{V}$. 
From that experimental data it seems that the dissipation rate does
increase with increased applied voltage rather than remaining constant:
we attribute this increase to dissipative channels not captured within
our approach. We also note that the current that flows in our thought
experiment is the same current defined in Ref.~\onlinecite{Balents2001}, 
but that the calculations
in that paper neglect the $\mathcal{O}(1/S)$ contribution to the current
that we calculate.

Quantitative comparison between theory and
experiment is complicated by the fact that the area of the sample over
which tunnelling takes place remains an open question\cite{Jack2004}. 
It may depend on the experimental geometry and on spatial fluctations
in the tunnelling amplitude. Our
calculation gives a current density per unit area of bilayer: in order
to match this data with experiments we must assume tunnelling taking
place over an area around $50\mu\mathrm{m}^2$, which is signicantly smaller
than the total area, so consistent with tunnelling taking place over
localised regions of the bilayer.

The calculation leading to equation~(\ref{equ:I_asp}) may be 
straightforwardly generalised to the
case in which an Aharonov-Bohm flux is introduced in the plane of the
bilayer. There is a momentum scale associated with the new flux, given by
$Q=(eB_y d/\hbar c)^{1/2}$ 
where $d$ is the spacing between the two quantum wells, and takes
a value around $30\mathrm{nm}$.  
With an appropriate gauge choice, the effect of the flux
is to introduce a spatial dependence to the tunnelling term in the 
Hamiltonian: $\dsas m^x \to \dsas (m^x \cos Qx + m^y\sin Qx)$. 

\begin{figure}
\epsfig{file=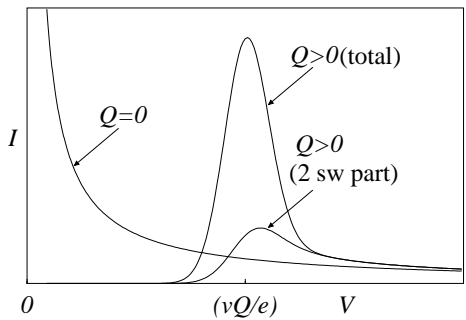,width=0.7\columnwidth}
\caption{Contributions to current at finite $Q$}
\label{fig:I_finiteQ}
\end{figure}

The result is that a new dissipative process appears in which a single
quantum of the $q=0$ mode decays into a single spinwave with momentum $Q$.
This leads to a contribution to the current at order $S^0$: in other words,
even the purely classical spin system has a decaying oscillation
in this case. Evaluating the leading two terms in the $(1/S)$ expansion,
we arrive at 
\begin{eqnarray}
I_{\theta>\theta_0,Q} &=&
\frac{D \dsasr^2}{16\pi\hbar l_B^2 } \times
\Big[  2\pi V^{-1} \delta(eV-vQ) +
 \nonumber\\
& & \phantom{hi} \;\frac{1}{2 \rho_E S} \frac{e}{\sqrt{(eV)^2-(vQ)^2}}
\Theta(eV-vQ) \Big]
\label{equ:I_finiteQ}
\end{eqnarray}
where $\Theta(x)$ is the step function.  The second term is the two
spinwave contribution that reduces to (\ref{equ:I_asp}) at $Q=0$.

The $\delta$-function in the first term indicates a resonance in the
classical spin system when the capacitative energy released on
transferring an electron across the bilayer matches the energy of the
collective mode with momentum $Q$. This infinitely narrow form of the
peak is clearly unphysical. This is because we are calculating
a transition rate to a discrete level, by specifying both the momentum
and the energy of the final one-spinwave state.  
The final state can itself decay at higher order in $1/S$ and so the delta
function will be broadened. 

Moreover, the delta function can be
broadened by disorder or by finite temperature. In
figure~\ref{fig:I_finiteQ} we show the $IV$ characteristic of
(\ref{equ:I_finiteQ}), where an uncertainty in the momenta of the
collective modes has been introduced to mimic the effect of weak
disorder. Quantitative calculation 
of the broadening introduced by these effects 
is beyond the scope of this paper: the degree broadening also controls
the peak height, which again frustrates a detailed comparison
with experiment. However, 
we emphasised in Ref.~\onlinecite{Jack2004} that the
asymmetric lineshape of this peak arises from the quantum corrections
that we have calculated: these are multi-spinwave decay channels.

\subsection{Damping of oscillations: $V<V_0$} 
\label{sec:qu_bsp}

We now turn to initial charging voltages $V<V_0$, in which the effect
of dissipation is to damp oscillations in the charge imbalance on the
bilayer. This behaviour was shown in the lower trace of
figure~\ref{fig:sz_t}. We note that this case is different from
$V>V_0$ in that our thought experiment does not result in a state with
a steady direct current. For that reason we cannot directly address
the $IV$ relation at these small biases: the strong feature in the
differential conductance observed
experimentally\cite{Spielman2000on} at small $V$ is beyond the scope
of this paper. 
However, the coherent oscillations of charge between the layers that are
predicted by the classical limit of the spin system are not observed in
experiments. The intrinsic dissipation rate that we calculate sets a
minimal decay rate for these oscillations. Adding disorder or inelastic   
scattering to the model might increase this decay rate: we consider only
the clean system in this paper, but the framework that we now describe
could be extended to consider this case.

The calculation of $\Gamma$ is slightly more complex than for $V<V_0$,
but proceeds along the same lines.  The result is a slow (power law)
decay of the oscillations predicted by the classical limit of the
pseudospin system.

In order to obtain a bosonic theory, we again use the
Holstein-Primakov representation. However, we may choose either the
$x$ axis or $z$ axis as the principal axis for our bosonic
theory. Choosing the $x$ axis is natural since that is the direction
of the mean magnetisation. On the other hand, choosing the $z$ axis
has the advantage that no interactions arise from the anisotropy term
in the Hamiltonian. We have checked that the results we present below
are independent of this choice. 

Let us consider first the spinwave spectrum $\omega_{\bm{q}}$.
The non-interacting part of the Hamiltonian for the spinwaves takes
the form
\begin{equation}
H_0 = \sum_{\bm{q}} \omega_{\bm{q}} \beta_{\bm{q}}^\dag \beta_{\bm{q}} 
\label{equ:swham}
\end{equation}
where $\beta_{\bm{q}}^\dag$ and $\beta_{\bm{q}}$ are the
creation and annihilation operators for a spinwave with momentum
$\bm{q}$ and
\begin{equation} 
\omega_{\bm{q}}=[(\dsas+\rho_E\gamma(\bm{q}))
(D+\dsas+\rho_E\gamma(\bm{q}))]^{1/2}. 
\label{equ:swdispersion}
\end{equation}
where $\gamma(\bm{q})$ was defined in section~\ref{sec:qu_asp}, above.  
Since we are treating the tunnelling non-perturbatively, the spinwaves 
now have an energy gap of $\dsw=[\dsas(D+\dsas)]^{1/2}$.

The non-interacting eigenspectrum has multiple branches, labelled by
the total number of spinwave quanta.  This should be contrasted with
the high-bias regime discussed in the previous section where there is
a qualitative difference between the $q=0$ and finite-$q$ modes and
the branches are labelled by the number of $q=0$ quanta. However, for
consistency, we continue to use the notation in which
$|N_0,\{\bm{q},\ldots\}\rangle$ describes a state with $N_0$ quanta in
the $q=0$ mode, as well as a set of collective excitations with finite
momenta. (We omit the set of finite momenta if it is empty.)

To set up the calculation for our thought experiment, 
the initial state of the system is chosen to be the ground state of the
bilayer in the presence of the gate voltage $V$. In terms of the
creation operator $\beta^\dag_{q=0}$ for the $q=0$ spinwave, 
this is approximated by a coherent state, which can be written as
\begin{equation}
|i\rangle_{V<V_0}=
 \mathcal{N} \exp\left[i\theta \sqrt{\frac{SL^2}{2c_0^2}} 
w_{q=0}\beta^\dag_{q=0}\right] | 0 \rangle
\label{equ:initial_state}
\end{equation}
where $w_{\bm{q}}=\sqrt{\omega_{\bm{q}}/(\dsas+\rho_E\gamma(\bm{q}))}$
is a coherence factor, and
$\mathcal{N}=\exp[-\theta^2 w_{q=0}^2 SL^2/ 4c_0^2]$ enforces
normalisation. 
The expectation of the occupancy of the $q=0$ mode in this
 state is
\begin{equation}
N_0 \equiv n_0 L^2 = (SL^2/2c_0^2) \theta^2 \sqrt{(D/\dsas)+1} 
\label{equ:boseocc}
\end{equation}
We observe that
the occupancy of the bosonic mode with $q=0$ is proportional 
both to the area of the system and to $S$: 
the bosons in this mode are Bose-condensed with areal density $n_0$. 
We will see below that interactions 
within the condensate must be treated carefully. 

We can now discuss the Hamiltonian for our calculation in this low-bias regime.
Our perturbative parameter for $V<V_0$ is $(n_0/S)$. It is
independent of $S$ and does not vanish in the classical limit.
We will calculate the dissipation rate $\Gamma$ to leading order in
this parameter.  We give the full details of the Hamiltonian in
the Appendix. It can be written in the form
\begin{equation}
H = H_0 + H_\mathrm{int}^c + H_\mathrm{diss}
\label{equ:lowbiasham}
\end{equation}
$H_\mathrm{int}^c$ describes the interactions within the condensate:
\begin{equation}
H_\mathrm{int}^c =
\frac{c_0^2}{L^2 S} \left[
\gamma_{4d} \beta_0^\dag \beta_0^3 +
\gamma_{4e} \beta_0^4 + \mathrm{h.c.} \right]
\label{equ:hamintc}
\end{equation}
and $H_\mathrm{diss}$ contains dissipative processes 
involving finite-momentum spinwaves
\begin{eqnarray}
  H_\mathrm{diss} &=& \frac{c_0^2}{L^2 S}
\sum_{\bm{q}\neq 0} \Big[ \frac{\gamma_{4a}}{2} \beta^\dag_{\bm{q}}
\beta^\dag_{-\bm{q}} \beta_0^2 + \frac{\gamma_{4b}}{2}
\beta^\dag_{\bm{q}} \beta^\dag_{-\bm{q}} \beta^\dag_0 \beta_0
\nonumber \\ & &
+ \gamma_{4c} (\beta^\dag_{\bm{q}} \beta_{\bm{q}}
+\beta^\dag_{-\bm{q}} \beta_{-\bm{q}}) \beta_0 \beta_0 + \textrm{h.c.}
\Big]
\nonumber \\ & &
+ \frac{c_0^4}{L^4 S^2} \sum_{\bm{q}\neq 0} 
\gamma_6 \beta^\dag_{\bm{q}} \beta^\dag_{-\bm{q}} \beta_0^4
 + \dots
\label{equ:qu_gammas}
\end{eqnarray}
where we have included only terms that
affect the dissipation rate, $\Gamma$, to leading order in
$(1/S)$. The vertex factors, $\gamma$'s, are 
functions of $\bm{q}$ which depend on the
choice of principal axis in the Holstein-Primakoff
transformation. They are independent of $L$ and are finite in the
large-$S$ limit.

Before discussing the dissipative processes, we note that we have
isolated the interactions within the condensate in $H_\mathrm{int}^c$
because these processes do not directly contribute to the dissipation of
energy from the $q=0$ modes into finite-momentum spinwaves. However,
at first sight, they do appear to contribute to the decay of our initial
state. Moreover, the rate for the process is apparently divergent in
both the thermodynamic (large $L$) and classical (large $S$) limits!
For instance, for the process $\gamma_{4e}$, 
Fermi's Golden Rule would give a transition amplitude of
\begin{equation}
\langle N_0-4 | H_\mathrm{int}^c | N_0 \rangle = (L^2 S) \gamma_{4e}
(n_0/S)^2 c_0^2 + \dots
\end{equation}
This seems to suggest that the condensate with $N_0$ $q=0$
modes decays immediately without the generation of finite-$q$ modes. This
divergence of the Fermi's Golden Rule amplitude 
originates from the macroscopic occupation of the $q=0$
mode. 

In fact, these condensate interactions do not cause decay
within the condensate. Instead, they contribute to a coherent
non-linear evolution of \emph{all} the condensed $q=0$ bosons: this
effect is beyond the perturbative framework of Fermi's Golden Rule. We
will see in the next section that this evolution is part of the
semiclassical spin dynamics. To avoid divergences, we need to modify
the interaction-picture operators to take into account the condensate
interactions: $O(t)=\exp[i \tilde{H} t] O \exp[-i\tilde{H} t]$ where
$\tilde H = H_0+H_\mathrm{int}^c$ contains the non-quadratic
interaction terms in $\beta_{q=0}$. This removes the divergent terms
from the Fermi's Golden Rule calculation, but we pay the price of a
non-trivial time dependence for the interaction-picture operators. For
details of the modifications, including some comments on its physical
interpretation, see appendix~\ref{app:mod_int}.

The effect of the non-linear evolution of the $q=0$ mode is
illustrated by the annihilation operator $\beta_{q=0}(t)$ for the
$q=0$ mode. For example, we have:
\begin{eqnarray}
\langle N_0-3 | \beta_{q=0}(t) | N_0 \rangle & = & 
  \frac{c_0^2}{L^2 S} \frac{\gamma_{4d}}{2\dsw} 
\langle N_0-3 | \beta_{q=0}^3 | N_0 \rangle 
\nonumber \\ & &
  \times ( e^{-3i\dsw t/\hbar} - e^{-i\dsw t/\hbar} )
\label{equ:nonlinearmode}
\end{eqnarray}
Thus $\beta_{q=0}(t)$ has a finite amplitude for destroying more than
one particle. (There is an analogous amplitude for particle creation
as well.) 
As a result of this non-trivial time dependence, the $\gamma_{4a}$
term in $H_\mathrm{diss}$ may now contribute to the decay of the initial state,
giving rise to a contribution to the dissipation rate proportional to
$\gamma_{4a}\gamma_{4d}$.

We now have all the ingredients required to calculate the dissipation
rate. We note that the dissipative part, $H_\mathrm{diss}$, contains only
terms of even order in the bosonic operators. Together with
energy-momentum conservation, we can see that the simplest transition
in which energy is transferred out of the $q=0$ mode involves the
annihilation of four quanta of that mode, combined with the creation
of a pair of spinwaves with equal and opposite momenta:
\begin{equation}
  |N_0\rangle \rightarrow |N_0-4, \{\bm{q},-\bm{q}\}\rangle
\end{equation}
where the momentum $\bm{q}$ has to satisfy energy conservation:
$4\dsw = 2\epsilon_{\bm{q}}$.
This kinematic constraint leads to an intrinsic dissipation
which depends on the fourth power of the density of spinwaves. At low
bias, this density is small and so the intrinsic dissipation is rather weak.

\begin{figure}
\epsfig{file=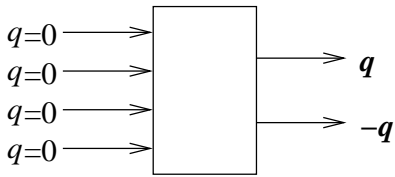,width=0.6\columnwidth}
\caption{
Schematic decay process for $V<V_0$. Energy conservation
requires $2\omega_{q}=4\dsw$}
\label{fig:diag_bsp}
\end{figure}

As expected from kinematics, the
dominant processes contributing to dissipation are of the form
shown in figure~\ref{fig:diag_bsp}. The modified interaction picture
yields an effective matrix element for the transition, given by
\begin{equation}
\gamma_{fi} = \dsw^{-1} (n_0 c_0^2/S)^2 (\gamma_{4b}\gamma_{4e} +
\gamma_{4a}\gamma_{4c}- \gamma_{4a}\gamma_{4d} - \dsw \gamma_6)
\label{equ:gamma_fi}
\end{equation} 
where the coefficients $(\gamma_{4a},\gamma_{4b},\dots)$ were
defined in equation (\ref{equ:qu_gammas}) as the vertex factors 
for various processes involving spin wave quanta.  The
different terms in $\gamma_{fi}$ correspond to different coherent
routes for destroying 4 $q=0$ spin wave quanta and generating a pair of
finite-$q$ spin waves.  This can be achieved with a direct vertex
($\gamma_6$), or through an intermediate virtual state $|N_0-2,\{\bm{q},-\bm{q}\}\rangle$ ($\gamma_{4a}$ followed by $\gamma_{4c}$). 
The $\gamma_{4a}$ term in $H_\mathrm{diss}$ also contributes to the
decay through the non-linear component of $\beta_{q=0}$ from
(\ref{equ:nonlinearmode}) giving the term $\gamma_{4a}\gamma_{4d}$.
The $\gamma_{4b}\gamma_{4e}$ term arises in a similar way.

Substituting the relevant coefficients into (\ref{equ:gamma_fi}), and
denoting the (small) 
number $(\dsas/D)$ by $x$, the result is
\begin{equation}
\gamma_{fi} = \dsw (n_0 c_0^2/S)^2 \frac{ 3 - 4x - 
2x \sqrt{1+16x+16x^2} - 8x^2}{128(1+x)^3}
\label{equ:gamma_fi_x}
\end{equation}
We have checked that this quantity is independent of 
the choice of quantisation axis. 
The density of states for spin wave pairs,
in which the spin waves have equal and opposite momentum, and each
has energy $\varepsilon$ is
\begin{equation}
g(\epsilon) = \frac{\varepsilon}{4\pi\rho_E c_0^2}
\frac{1}{\sqrt{D^2+4\varepsilon^2}}
\end{equation}
The dissipation rate is now simply given by
$\Gamma=8\pi L^2\dsw \gamma_{fi}^2 g(2\dsw)/\hbar$, and we arrive at
our result for the dissipation rate in the low-bias regime ($V<V_0$):
\begin{equation}
\Gamma_{V<V_0} = L^2
\left(\frac{V}{V_0}\right)^8 
 \frac{9\hbar\dsw^2\dsas^3}{512\pi l_B^2\rho_E D^2} [1+\mathcal{O}(\dsas/D)]
\label{equ:Gamma_bsp}
\end{equation} 
We can express this as the rate at which the number of $q=0$ modes is
decreasing: $dN_0/dt = - \Gamma(V) / 4\dsw$.  However, $N_0$ is
related to the instantaneous bias $V$ by (\ref{equ:boseocc}).  The
instantaneous bias is in turn related to the $z$-magnetisation by
(\ref{equ:tiltangle}): $V \propto m_z \sim \theta$ at low bias.
In terms of the tilt angle, we see that $d\theta^2/dt \propto -\theta^8$.
Hence the amplitude, $A$, of the
oscillating charge imbalance on the bilayer decays slowly in time,
according to
\begin{equation}
A(t) \sim 1/t^{1/6} 
\end{equation}

This is weak dissipation, but we 
believe that it is the first microscopic calculation of an intrinsic 
dissipation rate in these systems. The existence of such mechanisms is an 
important issue because
coherent oscillations of the charge are not observed in physical bilayer 
systems. Introducing disorder or inelastic scattering 
from topological excitations might relax the kinematic constraint that 
lead to the weak dissipation. Simulations\cite{Fertig2003} indicate that these 
effects are important at small energies: the framework of the modified 
interaction
picture may be a useful framework in which to investigate these effects.

Equations (\ref{equ:I_asp}), (\ref{equ:I_finiteQ}) and 
(\ref{equ:Gamma_bsp}) represent the main results of this section.
We have calculated dissipation rates in the three regimes of
our thought experiment that are most relevant to experimental systems.
As we commented in Ref.~\onlinecite{Jack2004}, 
the results agree qualitatively with experiment;
more quantitative agreement may require the taking into account both
disorder, and the effects of finite temperature.

\section{Classical origin of the dissipation}
\label{sec:cl}

In this section we show that the dissipation mechanisms in the bilayer
have their origins in dynamical instabilities of the classical
spin system with a Hamiltonian given by (\ref{equ:H_spins}). 
We consider the case $V<V_0$, and emphasise the strong links
between the calculation of the strength of the classical resonances and
the quantum calculation above. We emphasised in the previous section
that an initially coherent oscillation remains coherent forever in the
classical system. However, we show here that there are instabilities
of the classical dynamics whereby an initially very small spatial modulation 
of the spin field will grow resonantly, leading to dissipation out
of the coherent oscillation. In the quantum system, zero point fluctuations
are sufficient to seed this instability: this is the cause of the 
dissipation rates calculated in the previous section.

The classical equations of motion of the spin system are described by
the Landau-Lifshitz equation:
\begin{equation}
\hbar \frac{\mathrm{d}}{\mathrm{d}t} S_i^a = 
\epsilon^{abc} \frac{\partial H}{\partial S_i^b} S_i^c
\end{equation}
where $\vec{S}_i = (S^1_i,S^2_i,S^3_i)$ is the spin on site $i$
and $\epsilon^{abc}$ is the completely antisymmetric tensor.
We parametrise the magnetisation in terms of a mean value and a weak spatial
modulation : $\vec{m}_j = \vec{m}_0  + ( \vec{m}_{\bm{q}} e^{-i\bm{q}\cdot\bm{r}_j} + 
\hbox{c.c.})$. Assuming that the spatial modulation is small compared to the 
mean 
value, we may linearise the equation of motion for $\vec{m}_0$, arriving at
\begin{eqnarray}
(\hbar S) \frac{\mathrm{d}\vec{m}_0}{\mathrm{d}t} & = &  
  \vec{m}_0 \times \left[ \dsas \vec{e}_x - D \vec{e}_z m_0^z \right] 
\label{equ:m0_t}
 \\
(\hbar S) \frac{\mathrm{d}\vec{m}_{\bm{q}}}{\mathrm{d}t} & = &
  \vec{m}_{\bm{q}}
      \times \left[ \dsas\vec{e_x} - D \vec{e_z} m_0^z \right] \nonumber \\ & &
  + \vec{m}_0 \times \left[ - D \vec{e_z} m_{\bm{q}}^z
  - \rho_E\gamma(\bm{q})\vec{m}_{\bm{q}} \right]
\end{eqnarray}
where $\vec{e}_{x,y,z}$ are the unit vectors in the $x$,$y$,$z$-directions in 
spin space.

Let us consider first the time evolution of the uniform mode $\vec
m_0(t)$. Suppose we have an
initial state with a small component in the $S_z$-direction:
$\vec{m}_0^z(t=0)= (\sqrt{1-z_0^2},0,z_0)$ with $z_0\ll 1$.  We will
use $z_0$ as our perturbative parameter.
(Compare with the quantum
calculation of section~\ref{sec:qu_bsp} where the perturbative
parameter was the energy in the $q=0$ mode.)

At very small amplitudes, the driving mode is
harmonic: the spin precesses along elliptical trajectories with
frequency $\dsw/\hbar$. The perturbation expansion over $z_0$ is
straightforward, and is equivalent to solving for the motion of the
operator $\beta_{q=0}(t)$ in the modified interaction picture (see
appendix~\ref{app:mod_int}). The time dependence of the $z$-component of the 
uniform mode $m_0^z$ is
\begin{eqnarray}
m_0^z(t) &=&
 z_0 \cos (\tilde \dsw t/\hbar) + \frac{D z_0^3}{64 \dsw^2}\times 
\nonumber \\ 
&& \left[  
  \cos (3 \tilde \dsw t/\hbar) -
  \cos (\tilde \dsw t/\hbar) \right] 
+ \mathcal{O}(z_0^5) 
\end{eqnarray} 
where $\tilde \dsw=\dsw[1-(Dz_0/4\dsw)^2+\mathcal{O}(z_0^4)]$. 
The other components of $\vec{m}_0$ are 
easily obtained from $\dsas m_0^y=(\hbar S) (\mathrm{d}m_0^z/\mathrm{d}t)$ 
and $(m_0^x)^2 = 1 - (m_0^y)^2 - (m_0^z)^2$.

Let us now turn to the spatial modulations $\vec{m}_{\bm{q}}(t)$.
It is simple to verify that $|\vec{m}_0|$ and
$(\vec{m}_0 \cdot \vec{m}_{\bm{q}})$ are constants of the motion. To
make this explicit, we write
\begin{equation}
\vec{m}_{\bm{q}} = p_{2\bm{q}} ( \vec{m}_0 \times \vec{e_z} )
           + p_{1\bm{q}} ( \vec{m}_0 \times (\vec{m_0} \times \vec{e_z} ))
\end{equation}
This is the equivalent of casting the spin system (with three components
per site) in terms of bosons (one complex or two real degrees of freedom)
\emph{via} the Holstein-Primakov representation.
In this representation, the linearised equation of motion for the
spatial modulation is:
\begin{equation}
(\hbar S)
\frac{\mathrm{d}}{\mathrm{d}t} 
\left( \begin{array}{c} p_{1\bm{q}} \\ p_{2\bm{q}} \end{array} \right) =
\left( \begin{array}{cc} M_{d\bm{q}} & M_{a\bm{q}} 
  \\ -M_{b\bm{q}} & M_{d\bm{q}} \end{array} \right)
\left( \begin{array}{c} p_{1\bm{q}} \\ p_{2\bm{q}} \end{array} \right)
\label{equ:mq_t}
\end{equation}
where $M_{d\bm{q}} = \dsas m_0^y m_0^z/(1-(m_0^z)^2)$, 
$M_{a\bm{q}}=\dsas m_0^x +
\rho_E\gamma(\bm{q})$ and $M_{b\bm{q}}=[\dsas m_0^x / (1-(m_0^z)^2)] +
\rho_E\gamma(\bm{q}) + D(1-(m_0^z)^2)$.  

Consider first the limit of small amplitude $z_0\rightarrow 0$.
Then, $M_d = 0$, $M_a=\dsas +
\rho_E\gamma(\bm{q})$ and $M_b=\rho_E\gamma(\bm{q}) + D$. 
We can see that each $q$-mode is harmonic with a natural frequency
$\omega_{\bm{q}}/\hbar =(M_{a\bm{q}} M_{d\bm{q}})^{1/2}/\hbar$ 
which is the same as the dispersion relation 
given in (\ref{equ:swdispersion}). These are the classical 
spinwaves of the system.

However, at larger amplitudes $z_0$, we cannot ignore the driving of these
spinwaves by the uniform mode. This will provide the energy
to amplify any small spatial modulations.
Substituting $\vec{m}_0(t)$ into (\ref{equ:mq_t}), we can see that
these matrix elements are periodic in time.
They  take the form
\begin{equation}
M^{ab}_{\bm{q}}(t) = \sum_n M^{ab}_{\bm{q},n}\; e^{(i/\hbar) n\tilde\dsw t}
\end{equation}
We may therefore find solutions of the Bloch form 
\begin{equation}
\left( \begin{array}{c} p_{1\bm{q}} \\ p_{2\bm{q}} \end{array} \right) = 
e^{ik(\bm{q})t}
  \left( \begin{array}{c} u_{1\bm{q}}(t) \\ u_{2\bm{q}}(t) \end{array} \right)
\end{equation}
where $u_1$ and $u_2$ are periodic in $t$ with period 
$(2\pi\hbar /\tilde\dsw)$.
The relation between the wavevector of the spatial modulation,
$\bm{q}$, and the Bloch wavenumber, $k$, is a complex band structure, in which
complex wavenumbers represent resonantly growing (or exponentially decaying)
solutions.
The relation $k(\bm{q})$ is obtained by solving, 
for any given spatial mode $\bm{q}$,
an equation of the form 
\begin{equation}
\det_{ab,mn} \left\{
[k(\bm{q})+2n\tilde\dsw]\delta_{m-n}\delta^{ab}+iM_{\bm{q},m-n}^{ab}
\right\}=0
\end{equation}
where the determinant is of an (infinite dimensional) matrix operating 
in the product space of Fourier components ($mn$) and
components ($ab$) of the vector $\vec{p}$. 

\begin{figure}
\epsfig{file=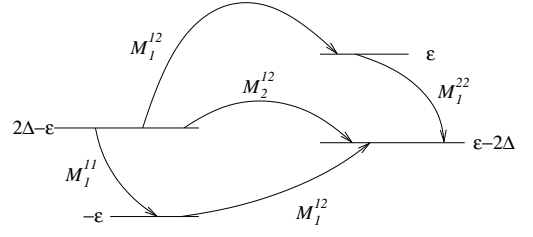,width=0.8\columnwidth}
\caption{`Energy level' picture of the classical calculation. Resonance
occurs when initial (left) and final (right) states become degenerate.}
\label{fig:cl_levels}
\end{figure}

This calculation closely resembles the calculation of the quantum
matrix element between initial and final states: see 
figure~\ref{fig:cl_levels}. The non-interacting ($z_0=0$) problem
defines energy levels which we may identify as initial and final states.
At finite $z_0$, we treat the interactions between the states
perturbatively: the resonance condition occurs when initial and final
states become close in energy (as in the quantum case).  

\begin{figure}
\epsfig{file=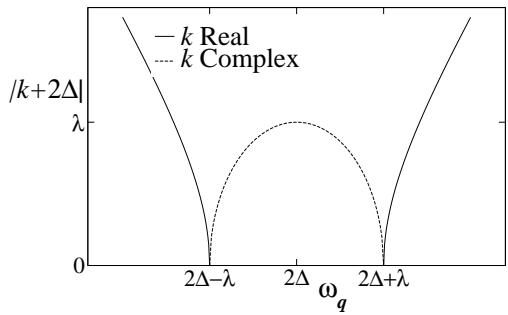,width=0.8\columnwidth}
\caption{Form of the complex band 
structure at the resonance ($\omega_{\bm{q}}=2\tilde\dsw$). 
We abbreviate the driving frequency $\tilde\dsw$ to $\Delta$.}
\label{fig:sc_gap}
\end{figure}

The resulting complex band structure for spatial wavevectors around
$\omega_{\bm{q}}= 2\dsw$ is shown in figure~\ref{fig:sc_gap}.  
There is a resonance at $2\omega_{\bm{q}} = 4\tilde\dsw$. At
the centre of the resonance, the imaginary part of the Bloch wavenumber, 
$\lambda$,
coincides exactly with the matrix element for the quantum decay process.  
To be precise,
\begin{eqnarray}
\lambda  & = & (\dsw/\hbar) (z_0/S)^4 (\dsw/2\dsas)^2 \times\nonumber \\
& & \qquad \frac{ 3 - 4x - 
2x \sqrt{1+16x+16x^2} - 8x^2}{128(1+x)^3}
\end{eqnarray}
where we have again abbreviated $(\dsas/D)=x$.
Identifying $\theta$ with $(z_0/S)$, and using equation~(\ref{equ:boseocc})
in conjunction with equation~(\ref{equ:gamma_fi_x}),
we see that
\begin{equation}
\lambda = \gamma_{fi}/\hbar
\label{equ:lambda}
\end{equation}

Equation~(\ref{equ:lambda}) is strong evidence that the quantum decay
process is linked to the instability of the classical system at the
same wavevector. (Note that even the 
non-trivial dependence on 
$\dsas/D$ contained in the factor $\gamma_{fi}$ is identical in both
cases).  

To make this correspondence closer,
we demonstrate that with an appropriate choice
of boundary conditions, the rate at which the classical oscillation decays is
approximately
equal to the rate given in (\ref{equ:Gamma_bsp}). 
An outline of this calculation is given in 
appendix~\ref{app:sc}. 
The boundary conditions of the classical calculation have been chosen
so as to mimic the quantum system. To set the boundary conditions,
the initial energy in the driven mode is of the order of zero point
energy
in that mode, and the initial phase difference between 
driven and driving mode is chosen such that the
energy can never drop below its zero point value.

\begin{figure}[hbt]
\epsfig{file=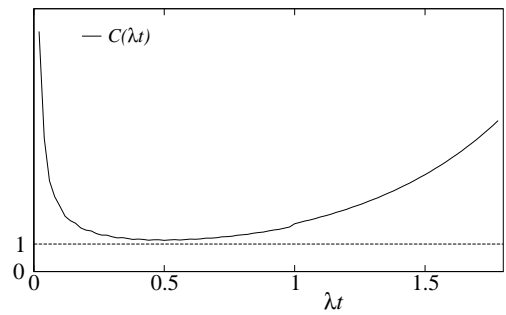,width=0.8\columnwidth}
\caption{
Classical (solid) and quantum (dotted) dissipation rates,
in units of the (constant) quantum rate. The time
dependence of the classical rate  follows the universal
function $\mathcal{C}(t)$: see (\ref{equ:sc_Ct}) and 
appendix~\ref{app:sc}.
}
\label{fig:Gammas_comp}
\end{figure}

The significance of the result (\ref{equ:Gamma_cl}) is that 
the dissipation rate in the classical system takes the universal
form:
\begin{equation}
\Gamma_\mathrm{cl} = 4\pi E(0) g_c(2\dsw) \lambda^2 \mathcal{C}(\lambda t)
\label{equ:sc_Ct}
\end{equation}
where $E(0)$ is the initial energy in the spinwave modes with frequency
$(2\dsw/\hbar)$, 
$g_c$ is the classical density of states at that frequency,
and the form of $\mathcal{C}(t)$ is shown in figure~\ref{fig:Gammas_comp}.  

Figure~\ref{fig:Gammas_comp} 
shows the similarities and 
differences between the classical and quantum dissipation rates.  
For the classical trace, we used
$E(0)=2\dsw$, which is of the order of the zero point energy in
the relevant spinwave mode. 
At short times ($\lambda t \ll 1$),
the classical system has a rapid transfer of energy
into modes far away from the resonance. Once this transient behaviour
is over, the quantum and classical rates are both approximately
constant in time at
intermediate times. The two results diverge again for long times
$\lambda t \gg 1$. However, 
the quantum calculation is non-degenerate perturbation
theory, and is therefore valid only up to times
$\lambda t \sim 1$ 
and so we do not expect the agreement to continue beyond this time scale.
We therefore emphasise that the form of the dissipation rate
calculated in appendix~\ref{app:sc} and plotted in
figure~\ref{fig:Gammas_comp} is a universal result for dissipation out of
a periodic oscillation into a bath of ``driven'' oscillators (as long
as the equations of motion of the driven oscillators may be
linearised). This may be regarded as the classical analogue
of Fermi's Golden Rule.

\section{Conclusion}
\label{sec:conc}

We have used a thought experiment on an isolated, clean system at 
zero temperature to investigate the link between tunnelling 
transport and dissipation in quantum Hall bilayers. The thought
experiment allows us to identify a crossover in the behaviour of the system
that may be underlying the small bias crossover in the tunnelling $IV$
characteristic. Our theory predicts an $IV$ characteristic
that is in qualitative agreement with experiment at large bias. 

In the very low bias
regime we have identified an intrinsic source of dissipation in the
system, and showed that its origin is in a dynamic instability of a
classical spin system. 

In the light of these results, two main questions remain. Firstly, 
we were unable to address the $IV$ characteristic at small biases, which
is the position of a prominent feature in the conductance. An extension
to the method so that a steady-state current can be forced through
the bilayer would be required if this issue is to be resolved.
Secondly, the effects of disorder on the model have not been considered, 
although this thought experiment and theoretical framework does seem 
appropriate for investigating that problem.

\begin{acknowledgments}
We gratefully acknowledge the financial support of EPSRC (in part 
through grant GR/S61263/01), and of the Royal Society.
\end{acknowledgments}

\begin{appendix}
\section{Modified interaction picture}
\label{app:mod_int}

In this appendix, we show how modifying the usual interaction picture
allows us to avoid divergences associated with interactions between 
quanta of the macroscopically occupied $q=0$ mode in the low-bias regime
(section \ref{sec:qu_bsp}). We begin by describing
our modifications, before commenting briefly on their physical interpretation.

For our calculation of the dissipation rate (\ref{equ:Gamma_def}) 
for the low-bias regime in section \ref{sec:qu_bsp}, we wish to evaluate matrix elements of the form
\begin{equation}
\mathcal{M}=
\langle 0,\{\bm{q},\bm{-q}\} | \beta_0^{N-m} e^{i H t} E_q e^{-i H t} 
(\beta_0^\dag)^N
| 0, \{ \} \rangle 
\label{equ:mat}
\end{equation}
where the operator $E_q$ measures the energy in the spinwave
mode with finite momentum $\bm{q}$ and 
$H$ is the Hamiltonian from (\ref{equ:lowbiasham}).
$N$ is a macroscopically large number and $m$ is a number of order unity.  For
convenience, we use the notation $\beta_0=\beta_{q=0}$ for the annihilation operator of the $q=0$ mode.

In the usual interaction picture, we would use time-dependent operators
$ O(t) = e^{iH_0t} O e^{-iH_0t}$ and the wavefunctions have a time dependence
of $|\psi(t)\rangle = e^{iH_0t} e^{-iHt} |\psi(0)\rangle$.
We would write
the matrix element in (\ref{equ:mat}) as: 
\begin{equation}
\langle 0,\{\bm{q},\bm{-q}\} | \beta_0^{N-m} e^{i H t} e^{-iH_0t} 
E_{q} e^{iH_0 t} e^{-i H t} (\beta_0^\dag)^N
| 0, \{ \} \rangle 
\end{equation}
where $H_0$ is
the quadratic Hamiltonian (\ref{equ:swham}) that commutes with $E_q$.
The product
$\hat{U}(t)=e^{iH_0t}e^{-iHt}$ may be written as a time-ordered exponential:
$\hat{U}(t)=T \exp[-i\int_0^t \mathrm{d}t' H_1(t')]$  which has a perturbative 
expansion in terms of
time integrals of the operator $H_1(t) = e^{iH_0t} (H-H_0) e^{-iH_0t}$.

Our modification to the interaction picture replaces the time
evolution $e^{-iH_0t}$ in the definition of the interaction picture
with $e^{-i\tilde{H}t}$ where $\tilde{H}$ includes the non-quadratic
terms from $H_\mathrm{int}^c$ (\ref{equ:hamintc}), but maintaining the
commutation relation $[E_q,\tilde{H}]=0$ for all modes except
$q=0$. Thus $\tilde{H}$ includes all terms that are functions only of
$(\beta_0,\beta^\dag_0)$: we have
\begin{eqnarray}
\tilde{H} & = & \sum_{\bm{q}} \omega_q \beta^\dag_{\bm{q}} \beta_{\bm{q}} +
H_\mathrm{int}^c + \dots
\nonumber \\ H_\mathrm{int}^c&=&
\frac{1}{L^2 S} \left[
\gamma_{4d} (\beta_0^\dag \beta_0^3 + \mathrm{h.c.}) +
\gamma_{4e} (\beta_0^4 + \mathrm{h.c.}) \right]
\label{equ:mi_H0}
\end{eqnarray}
where the omitted terms have coefficients that are 
$\mathcal{O}[(L^2 S)^{-2}]$. The
operator $H_1(t)$ must now be evaluated in a perturbative series
before the expansion of the exponential in $\hat{U}(t)$. This procedure
requires the (non-trivial) time dependence of the operator $\beta_0$
which is given (for $m\neq0$) by
\begin{eqnarray}
\langle N-m |  e^{i\tilde{H}t} \beta_0 e^{-i\tilde{H}t}  | N \rangle  =  
  \sqrt{N} \delta_{m-1} e^{-i\dsw t}  \nonumber \\ 
+ \left( e^{-im\dsw t} - e^{-i\dsw t} \right) \frac{\langle N-m |
[\beta_0,H_{int}^c] | N\rangle}{m\dsw}
 \nonumber \\ 
+ \sqrt{N}\mathcal{O}(N/L^2S)^2
\end{eqnarray}
These matrix elements converge even
in the classical (or thermodynamic) limit, in which $N\to\infty$ but
$NS^{-1}$ and $NL^{-2}$ are constants.

The purpose of this calculation is to evaluate a decay rate into a
continuum of final states. To interpret the modified interaction
picture in this light,
observe that we can write the matrix element, $\mathcal{M}$ in the form 
\begin{equation}
\mathcal{M}=\sum_{|f\rangle} 
\langle f | E_q | f \rangle \left| \langle f | e^{-iHt} | i \rangle \right|^2
\end{equation}
where $|i\rangle$ and $|f\rangle$ are initial and final states. Then,
writing $|f\rangle = e^{i\tilde{H} t}|g\rangle$, 
the fact that $\tilde{H}$ and $E_q$ commute
means that 
\begin{equation}
\mathcal{M}=\sum_{|g\rangle}
\langle g | E_q | g \rangle \left| \langle g | U(t) | i \rangle \right|^2
\label{equ:mi_M}
\end{equation}
Note that the state $|g\rangle=e^{-i\tilde{H} t}|f\rangle $ 
is obtained from $|f\rangle$ by
evolving the state $|f\rangle$ in parallel with the initial state,
taking into account the evolution of the condensate itself, but ignoring
interactions that involve modes with finite wavevectors.
This does not affect our results because the quantity
of interest, $E_q$, is independent of the occupancy of the condensate:
formally $\langle g | E_q | g \rangle=\langle f | E_q | f \rangle$.
In other words, this is a unitary change of basis for our subspace
of final states of the form $|N,\{\bm{q},\bm{q'},\dots\}\rangle$.

The advantage of using the $|g\rangle$ basis is that it allows us to
use conventional perturbation theory for the evolution operator
$U(t)$. Otherwise, we find divergences when evaluating matrix elements
to individual final states in the $|f\rangle$ basis. Physically, these
divergences arise from the fact that the particles in the condensate
have evolved in time and so the initial condensate $|N_0\rangle$ has
become strongly admixed with states with different numbers of $q=0$
modes: $|N_0+m,\{\bm{q},-\bm{q}\}\rangle$. 
The $|g\rangle$ basis tracks this evolution and is therefore the
natural basis for the calculation.

Equations (\ref{equ:mi_H0}) and (\ref{equ:mi_M}) 
define a convergent perturbative 
expansion for the dissipation rate $\Gamma$, defined in (\ref{equ:Gamma_def}): 
this leads to equation (\ref{equ:Gamma_bsp}).

\section{Dissipation rates in classical systems}
\label{app:sc}

In this appendix we discuss the extent to which a broad class of 
quantum problems with solutions based on Fermi's Golden Rule may be addressed
within a classical framework. We investigate the decay of a coherent 
oscillation by dissipation into an environment modelled by a `bath' of
harmonic oscillators. The only quantum ingredient will be the zero-point
fluctuations of this bath.

To illustrate the point we use a simple model system with a Hamiltonian
of the form
\begin{equation}
H = \frac{1}{2}(p_y^2 + \Omega^2 y^2) + \sum_n \left[
   \frac{1}{2} (p_{xn}^2 + \omega_n^2 x_n^2 ) + 4\lambda_n\Omega x_n^2 y^2
\right]
\label{equ:H_toy}
\end{equation}
where the coordinate $y$ describes the coherent (driving) oscillation
and the coordinates $x_n$ describe the bath of driven oscillators.
We work perturbatively around $\lambda=0$ which is the limit of independent
oscillators in which the coherent oscillation remains coherent for ever.
The energy in the driven oscillator is initially much greater than all
other energy scales, so $y\simeq \cos\Omega t$, and the $x_n$ obey
Mathieu's equation\cite{BenderBook}  for small $x_n$.
\begin{equation}
(\partial_t^2 + \omega_n^2 + \lambda \sin 2\Omega t) x_n \simeq 0
\label{equ:mathieu}
\end{equation}
For each $n$, solutions can be written in the Bloch form 
$x_n(t) = e^{ikt} u_n(t)$
where $u_n(t)$ has a period of $2\pi/\Omega$.

The interesting behaviour occurs when $(\omega_n/\Omega)$ is very close
to an integer. The Bloch wavenumber $k$ becomes complex and gaps open up.
Near $\omega_n=\Omega$, it can be shown that
\begin{equation}
(k+\Omega)^2 = (\omega_n-\Omega)^2 - \lambda_n^2
\end{equation}
Thus there is a gap at the edge of the Brillouin zone:
$\mathrm{Re}(k)=-\Omega$.  For $|\omega_n-\Omega|>\lambda_n$, 
$k$ is real and the solution for a given $n$-mode is
\begin{equation}
x_n^{(r)} = \mathrm{Re}\left[ Re^{i\phi} \left(
(k+\omega_n-2\Omega_n) e^{i(\Omega-k)t} - \lambda_n e^{i(k-\Omega)t} \right)
\right]
\end{equation}
where $R$ and $\phi$ are determined by boundary conditions. 
For $|\omega_n-\Omega|<\lambda_n$, $k$ is complex. Writing
$\kappa=i(k-\Omega)$ (which
is real) and $2\theta=\arg(i\kappa-\omega_n+\Omega)$, the solution is
\begin{equation}
x^{(i)}_n = A e^{\kappa t} \cos(\Omega t+\theta) - 
      B e^{-\kappa t} \cos(\Omega t - \theta)
\label{equ:sc_xi}
\end{equation}
where $A$ and $B$ are determined by initial conditions.

In our quantum calculation based on Fermi's Golden Rule, we
calculated the energy transferred to the driven oscillators, using for
the energy of each mode the energy given by the non-interaction
Hamiltonian. The analogous quantity here would be the dissipation rate
\begin{equation}
\Gamma_\mathrm{cl} = \partial_t \sum_n (1/2) ( p_{xn}^2 + \omega_n^2
x_n^2).
\end{equation}
Using the solutions above, 
the energy in each driven mode with real $k$ is
\begin{equation}
E_n^{(r)} = E(0) \frac{w_n - \lambda_n\cos[2(k+\Omega)t+2\phi']}{w_n -
\lambda_n\cos 2\phi'}
\label{equ:sc_er}
\end{equation}
with $w_n=(\lambda_n^2+|k+\Omega|^2)^{1/2}$. The angle $\phi'$ is equal 
to $\phi$ for $\omega>\Omega$: if $\omega<\Omega$ then
$\phi'=\phi+(\pi/2)$. Similarly, if $k$ is complex then we have
\begin{equation}
E_n^{(i)} = E(0) \frac{\lambda_n(\cosh2\kappa t+\sinh 2\kappa t
\sin\phi'')-w_n\cos\phi''}{\lambda_n -
w_n\cos \phi''}
\label{equ:sc_ei}
\end{equation}
where the phase angle $\phi''$ depends on the ratio of $A$ and $B$
in equation~(\ref{equ:sc_xi}).

For correspondence between the classical and quantum calculations, the initial
energy in the driven oscillators should be of the order of the zero point
value of the quantum calculation. Further, we choose the relative phase
between driven and driving oscillators such that the energy in the classical
oscillators can never be reduced below the same zero point value (since that
would violate the uncertainty principle). 

Finally the result is that 
\begin{equation}
\Gamma_\mathrm{cl} = E(0) \lambda^2 g_c(\Omega) \mathcal{C}(\lambda t)
\label{equ:Gamma_cl}
\end{equation}
where $g_c(\omega)$ is the classical density of states : 
$\sum_n \to \int \mathrm{d}\omega \, g_c(\omega)$ in the limit in the 
frequencies in the oscillator bath form a continuum. The universal
function $\mathcal{C}(\lambda t)$ was plotted in
figure~\ref{fig:Gammas_comp}. It takes the form
\begin{equation}
\mathcal{C}(x) \simeq \left\{ \begin{array}{ll} (1/2\pi x)-(1/4) & x\ll 1 \\
1 + \mathcal{O}(x) & x>1\end{array} \right.
\end{equation}
The diverging rate at small $x$ is cut off by the finite width of
the oscillator bath. At small times there is a rapid transfer of a small 
amount of energy into modes far from the resonance.

The rate calculated in (\ref{equ:Gamma_cl})
 is compared with the dissipation rate for a quantum calculation
on the same system in figure~\ref{fig:Gammas_comp}. 
The agreement breaks down at long times 
since the classical calculation corresponds to degenerate perturbation
theory for states near the resonance, and is able to capture the resonant
growth of the energy. At small times there is a transient effect that is
unique to the classical system. However, at intermediate times then choosing
$E_{0}=\hbar\omega_n$ gives good agreement between quantum and classical rates. 
This is in accordance with the correspondence principle, as expected for
systems with linear equations of motion. 

The applicability of this treatment to more complicated models than
that of equation (\ref{equ:H_toy}) 
is not obvious at first sight. However, the
forms for the energy in the driven modes, (\ref{equ:sc_er}) and
(\ref{equ:sc_ei}), are in fact universal\cite{JackThesis} for 
problems in which the equation of motion of the driven modes is
linearisable (that is, problems in which the periodic forcing  
multiplies a coordinate as in (\ref{equ:mathieu}), 
so that the forcing vanishes as the
energy in the driven oscillator gets small). In this general problem the 
parameter $\lambda$ in (\ref{equ:Gamma_cl}) 
is equal to the maximal value of the imaginary 
part of the Bloch wavenumber. Neglecting the effect of the initial transient
on the upper cutoff, (\ref{equ:Gamma_cl}) 
is then a universal function of $\lambda t$,
so the dissipation rate associated with resonance is completely characterised
by the $\lambda$ and the density of driven oscillators $g_c(\omega)$.
This is the general result applicable to the spin system of 
equation~(\ref{equ:H_spins}), and shows that the quantum decay channels
have corresponding instabilities in the classical dynamics. 

\end{appendix}

\end{document}